\documentclass[showpacs,twocolumn,prl,floatfix]{revtex4}
\usepackage{graphicx}
\usepackage{times}
\usepackage{bm}
\usepackage{amsmath}
\DeclareMathOperator{\var}{var}

\begin{document}

\date{August 2002}
\title{Band-center anomaly of the conductance distribution in one-dimensional 
Anderson localization}

\author{H. Schomerus}
\author{M. Titov}
\affiliation{Max-Planck-Institut f\"ur Physik komplexer Systeme,
N\"othnitzer Str. 38, 01187 Dresden, Germany
}

\begin{abstract}
We analyze the conductance distribution function
in the one-dimensional Anderson model of localization,
for arbitrary energy.
For energy at the band center
the distribution function
deviates from the universal form assumed in single-parameter scaling theory.
A direct link to the break-down of the random-phase approximation is
established.
Our findings are confirmed by a parameter-free comparison
to the results of numerical simulations.
\end{abstract}

\maketitle

The spatial localization of waves in a disordered potential
can be considered as the most dramatic  
effect of multiple coherent wave scattering \cite{sheng,berk}.
Due to systematic constructive interference
in some part of the medium  
the wave function is spatially confined  and decays
exponentially as one moves away from the localization center
\cite{Anderson,reviews}.
The localization length $l_{\rm loc}$
can be probed non-invasively from the decay of the
transmission coefficient (the dimensionless conductance
\cite{Landauer}) $g$, in terms of
the average
\begin{equation}
C_1\equiv \langle -\ln g\rangle = 2L/l_{\rm loc} + O(L^0)
\label{eq:c1}
\end{equation}
for system length $L\gtrsim l_{\rm loc}$ \cite{Borland}. 
Localization results in insulating behavior of
disordered solids at low temperatures \cite{Anderson,reviews},
and also can be realized in electromagnetic waveguides \cite{Genack},
where it is considered as an efficient feedback mechanism
for lasing in disordered active media \cite{Cao}.

One of the cornerstones of the theoretical understanding of
localization is the universal approach of
single-parameter scaling (SPS)
\cite{abrahams,anderson1980,shapiro}.
In this theory it is assumed that the complete
distribution function $P(g)$ of the conductance
can be parameterized by the single free parameter
$C_1$.
The dependence of $C_1$ [and hence of $P(g)$] on  $L$ is then found from solving a
scaling equation
$dC_1/d(\ln L)=\beta(C_1)$,
where the universal scaling function $\beta$ does not depend  on
$L$, nor on any microscopic parameter
(like the Fermi wavelength $\lambda_{\rm F}$, the transport mean free path
$l_{\rm tr}$,  or the lattice constant $a$).

The distribution function $P(g)$ is completely determined by the
cumulants
\begin{equation}
C_n\equiv  \langle\langle (-\ln g)^n\rangle \rangle,
\label{eq:2}
\end{equation}
which are obtained as the expansion coefficients
of  the generating function
\begin{equation}
\eta(\xi)=\ln\langle g^{-\xi} \rangle
=\sum_{n=1}^\infty C_n\frac{\xi^n}{n!}.
\label{eq:cumulgen}
\end{equation}
The first three cumulants are given by Eq.\ (\ref{eq:c1}) for $C_1$, $C_2=\var
\ln g$, and $C_3=\langle(\langle\ln g\rangle-\ln g)^3\rangle$.
The SPS hypothesis can then be phrased like this: {\em All cumulants
are universal functions of $C_1$.}
In the localized regime ($C_1\gg 1$), 
the universal SPS relations take the simple form \cite{anderson1980}
\begin{equation}
C_n/C_1=\delta_{1n}+2 \delta_{2n} +O(L^{-1}).
\label{ccond}
\end{equation}
These conditions are much more restrictive than the general upper bound
$C_n=O(L/l_{\rm loc})$ from the theory of
large-deviation statistics \cite{ott,ellis}:
SPS assumes a
lognormal distribution of $g$,
with the variance of $\ln g$ determined by the mean via
the universal relation $\var  \ln g = -2\langle \ln g\rangle$.
It is the violation of this relation which frequently is used 
to indicated the break-down of SPS theory
(see, e.g., Ref.\ \cite{Altshuler1,Altshuler2}).

In this paper we investigate $P(g)$ in
the most-studied and best-understood paradigm of
localization,
the one-dimensional Anderson model defined by the Schr\"odinger equation
\begin{equation}
\psi_{l-1}+\psi_{l+1}=(V_l-E)\psi_l
\label{eq:1}
\end{equation}
on a linear chain of $L$ sites (lattice constant $a=1$)
and a random potential
with $\langle V_l\rangle=0$ and
$\langle V_l V_m \rangle= 2D \delta_{lm}$.
The strength $D$ of the potential fluctuations 
is taken to be small. 
We analytically calculate the cumulants $C_n$
in the localized regime, with main focus on the energy region
$|E|\ll 1$ around the band center of the disorder-free system.
For $E=0$ we find the values 
\begin{equation}
C_2/C_1=2.094,\quad C_3/C_1=0.568.
\label{eq:cratios}
\end{equation}
The ratios $C_n/C_1$ with the higher cumulants also are finite.
Hence $P(g)$ complies with the restrictions
of large-deviation statistics, but
deviates from the special lognormal form
assumed in SPS theory (this form is restored for
$|E|\gtrsim D$).

The conditions for validity of SPS have been a constant subject of
intense debate \cite{shapiro,Altshuler1,Altshuler2}.
Originally, SPS was derived within the
random-phase approximation (RPA) for the scattering phase between
consecutive scattering events \cite{abrahams,anderson1980}.
In the Anderson model the RPA is known to fail around the 
energies $E=\pm 2$ (the band edges of the disorder-free system) 
\cite{lifshitz}, where $\lambda_{\rm F}\gtrsim l_{\rm tr}$. 
Indeed, the SPS relations (\ref{ccond}) are violated for all cumulants
when one comes close to the band edge
($2-|E| \lesssim D^{2/3}$)
\cite{finitetimelyap}, in coincidence with the expectations
\cite{lifshitz,Altshuler1,Altshuler2,Deych}.

The RPA is also known to break down for
the band-center case $E=0$ \cite{Stone}.
However, the only
consequence observed so far has been a weak anomaly in the
energy-dependence of $l_{\rm loc}$ (hence, also of $C_1$)
\cite{kappus,goldhirsch}, which differs at $E=0$
by about $9\%$ from the predictions
of perturbation theory \cite{Thouless1979}.
Surprisingly, the violation (\ref{eq:cratios})
of the SPS relations (\ref{ccond}) has not been noticed---quite
the contrary,
the relevance of the RPA for SPS recently has been contested 
\cite{Altshuler1,Altshuler2} within an  investigation of the  Lloyd
model, given by Eq. (\ref{eq:1})
with a Cauchy distribution
for the potential  \cite{lloyd,lifshitz}.
However, results obtained for the Lloyd model
are not conclusive  for the Anderson model and SPS,
because in the Lloyd model 
formally $D=\infty$ and one encounters the modified
universal relations $C_2/C_1=4\neq 2$, while
$l_{\rm loc}$ varies smoothly with energy even 
around $E=0$ \cite{lifshitz}.
Moreover, the higher cumulants have not been investigated.
In previous numerical studies, the violations may have passed unnoticed 
because the small deviation  of $C_2/C_1$ from the SPS value
probably was not considered to be significant,
and again the higher cumulants have not been investigated.
In this paper, we also will establish a direct link between SPS and RPA.

We now present the analytical calculation of the cumulants $C_n$
of $-\ln g$ in the vicinity of
the band-center energy $E=0$ of the Anderson model, Eq.\ (\ref{eq:1}).
As pointed out many years ago by Borland \cite{Borland}, the dimensionless 
conductance $g$ in the localized regime is statistically equivalent 
to $\psi_L^{-2}$, where $\psi_L$ is the solution of the 
Schr\"odinger equation (\ref{eq:1}) with generic initial conditions
$\psi_0$, $\psi_1=O(1)$.
Because $\lambda_{\rm F}\simeq 4 a$,
it is useful to introduce two slowly varying fields
$\phi(l)=\psi_{l}(-1)^{l/2}$ when $l$ is even,
$\chi(l)=\psi_{l}(-1)^{(l+1)/2}$ when $l$ is odd,
which can be considered as continuous
functions with Langevin equations
\begin{equation}
\frac{d \phi}{dL}=\frac{1}{2}(U-E)\chi,
\qquad \frac{d \chi}{dL}=\frac{1}{2}(W+E)\phi.
\label{eq:langevin}
\end{equation}
Here $U$ and $W$ independently fluctuate
with $\langle U\rangle=0$, $\langle U(L_1)U(L_2)\rangle=4D\delta(L_1-L_2)$,
and analogously for $W$.

In order to calculate the wave-function decay and its fluctuations
it is convenient to switch to the variables
\begin{equation}
u=\ln(\phi^2+\chi^2),\qquad
\sin\alpha=\left(\frac{\phi}{2\chi}
+\frac{\chi}{2\phi}\right)^{-1},
\end{equation}
which are symmetric in $\phi$ and $\chi$.
In the localized regime, $u=-\ln g$
characterizes the global decay of the wave function,
while the  variable $\alpha$
(parameterizing the local fluctuations) is identical to the
scattering phase of
the reflection amplitude $r=(\psi_{L-1}+i\psi_L)/(\psi_{L-1}-i\psi_L)$.
This parameterization allows us to draw a
direct relation between SPS and RPA: SPS will turn out to be valid
when $\alpha$ is uniformly distributed over $(0,2\pi)$.

The Langevin equations (\ref{eq:langevin}) now can be translated
into a Fokker-Planck equation for the joint distribution function
$P(u,\alpha;x)$.
For the sake of a compact presentation we use short-hand
notations for the functions
$s_\alpha=\sin \alpha$, $c_\alpha=\cos \alpha$,
and introduce the rescaled position $x=D L$,
as well as the rescaled energy $\varepsilon=E/D$.
The Fokker-Planck equation then takes the form
\begin{eqnarray}
&&\partial_{x}P(u,\alpha;x)=
\nonumber\\
&&
[{\cal L}_\alpha^2  +
\partial_u\left(s_\alpha^{2}\partial_u-c_\alpha^2
+2\partial_\alpha s_\alpha c_\alpha
\right)-\varepsilon\partial_\alpha ] P(u,\alpha;x),\qquad
\label{eq:fpua}
\end{eqnarray}
with the linear differential operator
${\cal L}_\alpha=\partial_\alpha (1+c_\alpha^2)^{1/2}$.

The behavior of $P(u,\alpha;x)$  
for large $x$ can be analyzed
by introducing into Eq.\ (\ref{eq:fpua}) the ansatz
\begin{equation}
P(u,\alpha;x)=\int_{-i\infty}^{+i\infty}\frac{d\xi}{2\pi i}
\sum_{k=0}^\infty \exp(\mu_k (\xi) x-\xi u)f_k(\xi,\alpha),
\label{eq:ansatz}
\end{equation}
where we require periodicity of $f_k(\xi,\alpha)$ in $\alpha$.
It then follows that the functions $f_k(\xi,\alpha)$
solve the eigenvalue equation
\begin{equation}
\mu_k f_k=\left[{\cal L}_\alpha^2-\varepsilon\partial_\alpha
+\xi(c_\alpha^2-2\partial_\alpha s_\alpha c_\alpha)
+\xi^2 s_\alpha^2
\right]f_k,
\label{eq:eval1}
\end{equation}
in which $\xi$ appears as a parameter and $\mu_k(\xi)$
is the $k$th eigenvalue (arranged in descending order).
In the vicinity of $\xi=0$, there is a finite gap between the largest
eigenvalue $\mu_0$ [which vanishes for $\xi=0$, because of the
normalization of $P(u,\alpha;x)$]
and $\mu_1$.
According to Eq.\ (\ref{eq:ansatz}),
the asymptotic behavior
of the distribution function $P(u,\alpha;x)$ for large $x$
hence is governed by
$\mu_0$, up to exponentially small corrections.
A formal calculation of the moments of $u$ (i.e., of $-\ln g$)
shows that 
the cumulant-generating function (\ref{eq:cumulgen}) is directly given by
$\eta(\xi)=x \mu_0(\xi)$. Hence, 
\begin{equation}
C_n= \mu^{(n)}n! D L,
\label{eq:mucum}
\end{equation}
where we expanded
$\mu_0(\xi)=\sum_{n=1}^\infty \mu^{(n)}\xi^n$
into a power series.

The expansion coefficients $\mu^{(n)}$ can be calculated 
recursively for increasing
order $n$ by solving the hierarchy of equations
\begin{eqnarray}
\sum_{k=0}^n
\mu^{(n-k)} f^{(k)}&=&
s_\alpha^2 f^{(n-2)}
+
(c_\alpha^2-2\partial_\alpha s_\alpha c_\alpha)f^{(n-1)}
\nonumber\\
&&{}+
{\cal L}_\alpha^2 f^{(n)}-\varepsilon
\partial_\alpha f^{(n)}
,
\label{eq:hier}
\end{eqnarray}
which results when one
introduces into Eq.\ (\ref{eq:eval1}) the power expansions for $\mu_0$
and for
$f_0(\xi,\alpha)=\sum_{n=0}^\infty f^{(n)}(\alpha)\xi^n$:
In each order $n$, we first integrate over $\alpha$ from
$0$ to $2\pi$, which eliminates $f^{(n)}$ and hence gives
$\mu^{(n)}$ in terms of the quantities $f^{(m)}$ and  $\mu^{(m)}$
with $m<n$. Afterwards $f^{(n)}$ can be obtained
from Eq.\ (\ref{eq:hier})
by two integrations.
The iteration is initiated for $n=0$ with $\mu^{(0)}=0$.
This completely solves the problem to calculate the cumulants $C_n$ in the
localized regime.

Let us illustrate the procedure for $E=0$.
To start the iteration we
consider Eq.\ (\ref{eq:hier}) with $n=0$, given by 
${\cal L}_\alpha^2 f^{(0)}=0$.
This differential equation
is solved by the normalized function
\begin{equation}
f^{(0)}(\alpha)=\frac{\sqrt{2\pi}}{\Gamma^2(1/4)\sqrt{1+\cos^2 \alpha} },
\label{eq:f0}
\end{equation}
which is identical to the
stationary limiting-distribution function 
$\lim_{x\to\infty}\int_{-\infty}^\infty du\, P(\alpha,u;x)$
of the variable $\alpha$. 

Now the next iteration. Equation (\ref{eq:hier}) with $n=1$ is given by
\begin{equation}
{\cal L}_\alpha^2 f^{(1)}(\alpha)
=
\left(\mu^{(1)}-c_\alpha^2
+2\partial_\alpha s_\alpha c_\alpha\right)f^{(0)}(\alpha).
\end{equation}
We first determine
\begin{equation}
\mu^{(1)}=\int_{0}^{2\pi} d\alpha\, c_\alpha^2 f^{(0)}(\alpha)
=4 \frac{\Gamma^2(3/4)}{\Gamma^2(1/4)}.
\label{eq:mu1}
\end{equation}
The prediction for the inverse localization length
\begin{equation}
l_{\rm loc}=  \Gamma^2(1/4)/[2 D \Gamma^2(3/4)],
\label{eq:loclen}
\end{equation}
obtained by combining  Eq.\ (\ref{eq:mu1}) with
 Eqs.\ (\ref{eq:c1}) and (\ref{eq:mucum}),
is identical to the result found in Refs.\ \cite{kappus,goldhirsch}.
Then we solve for
\begin{eqnarray}
&&f^{(1)}(\alpha)=(1+c_\alpha^2)^{-1/2}
\int_0^\alpha d\beta\, (1+c_\beta^2)^{-1/2}
\Big[
\nonumber \\
&&\quad{}
2s_\beta c_\beta f^{(0)}(\beta)
+\int_0^\beta d\gamma\,
(\mu^{(1)}-c_\gamma^2)f^{(0)}(\gamma)\Big]
.\quad
\end{eqnarray}
From the next iteration $n=2$
we obtain 
\begin{equation}
\mu^{(2)}=\int_{0}^{2\pi} d\alpha\,
[(c_\alpha^2-\mu^{(1)})f^{(1)}(\alpha)+s_\alpha^2f^{(0)}(\alpha)]
\end{equation}
and  also
$f^{(2)}(\alpha)$. Analogously we obtain $\mu^{(3)}$.
With Eq.\ (\ref{eq:mucum}), this is sufficient to
determine the
values for the first three cumulants
\begin{equation}
C_1=  0.4569\,D L,\quad
C_2=  0.9570\,D L,\quad
C_3=  0.2595\,D L.
\label{eq:c123}
\end{equation}
They correspond to the anomalous 
ratios given in Eq.\ (\ref{eq:cratios}).

\begin{figure}[t]
\includegraphics[width=\columnwidth]{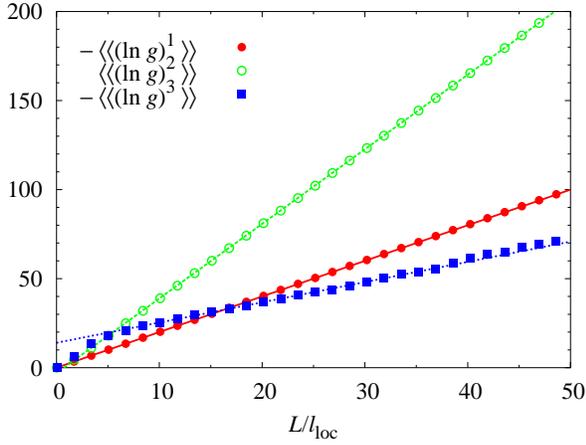}
\caption{First three cumulants $C_n=\langle\langle(-\ln g)^n\rangle\rangle$
for energy $E=0$ in the Anderson model (\ref{eq:1}) with $D=1/150$, 
as function of system length $L$.
The data points are the result of a numerical simulation.
The slopes of the straight lines follow the predictions of Eq.\ (\ref{eq:c123}).
The localization length $l_{\rm loc}$ is taken from Eq.\ (\ref{eq:loclen}).
}
\label{fig1}
\end{figure}

\begin{figure}[t]
\includegraphics[width=\columnwidth]{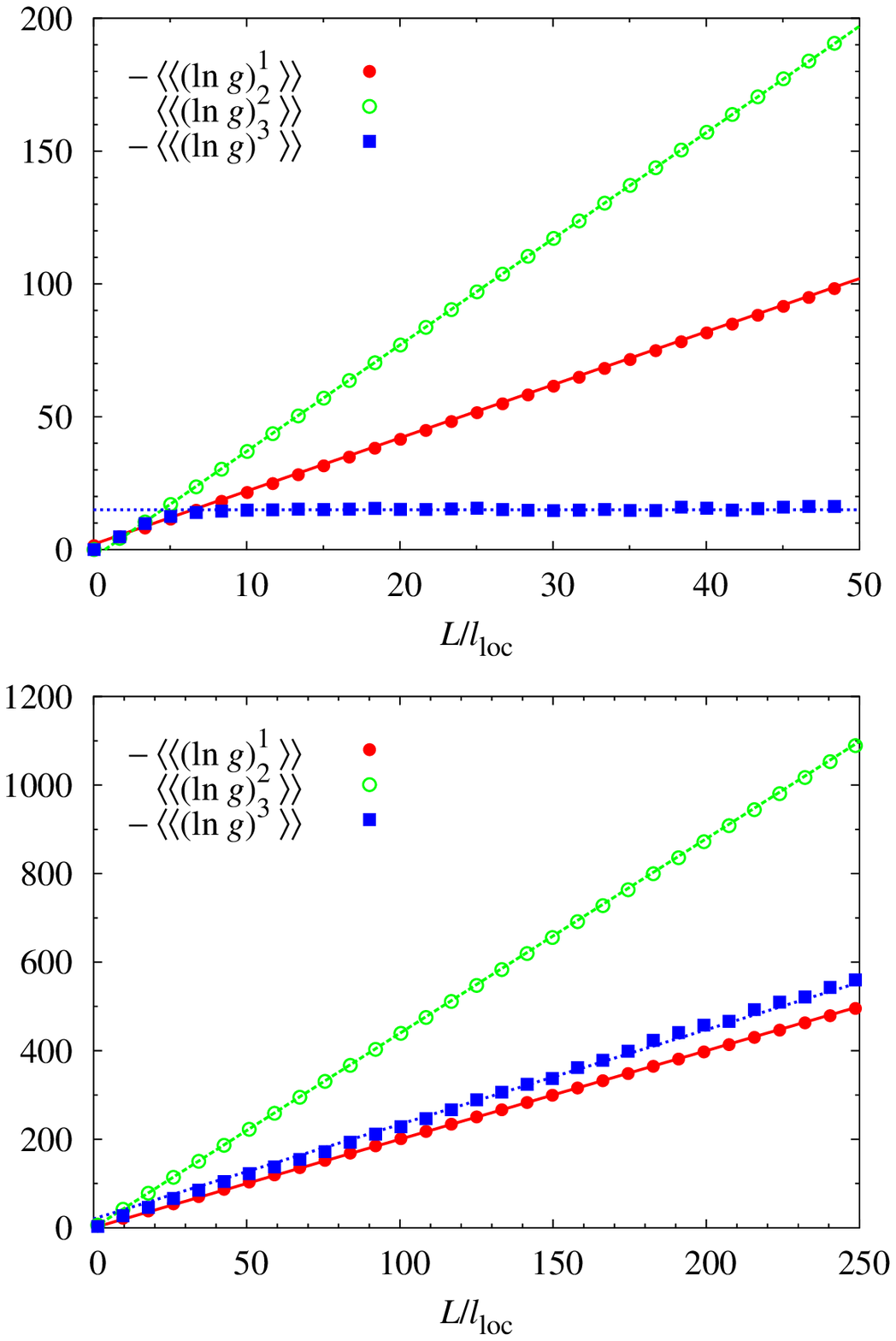}
\caption{Same as Fig.\ \ref{fig1}, but for 
energy $E=0.1$ (upper panel) and $E=2$ (lower panel).
The straight lines in the upper panel follow the predictions
of perturbation theory \cite{Thouless1979} and
single-parameter scaling \cite{anderson1980}.
The straight lines in the lower panel
are the predictions 
of Ref.\ \cite{finitetimelyap} (see text).
}
\label{fig2}
\end{figure}

The analysis of Eq.\ (\ref{eq:hier}) can be straightforwardly carried out
also for finite $E/D$.
For $E/D\gg 1$, the stationary limiting-distribution function
of $\alpha$ is given by $f^{(0)}(\alpha)=
1/(2\pi)$, corresponding to a completely random phase.
For $n=1$ we find the coefficient
$\mu^{(1)}=1/2$, and the perturbative result $l_{\rm loc}=4/D$
is recovered \cite{Thouless1979}.
In the next iteration we obtain $\mu^{(2)}=1/2$, while the higher coefficients
all vanish.  According to Eq.\ (\ref{eq:mucum}),
the SPS relations (\ref{ccond}) then are reestablished.

We have tested the predictions of the analytical theory
against the result
of a direct numerical computation of the conductance  $g$
for the Anderson model (\ref{eq:1}),
by recursively increasing the length of the wire \cite{jalabert}.
The potential $V_l$ was drawn independently for each site from a box
distribution with uniform  probability $1/\sqrt{24 D}$ over the interval
$[-\sqrt{6D}, \sqrt{6D}]$. The data shown in the plots was obtained
for $D=1/150$
(identical results are obtained for a Gaussian distribution
with the same variance $D$).
The cumulants were determined by averaging over $10^7$ disorder
realizations.

The result of this computation for the first three cumulants
and $E=0$
is shown in Fig.\ \ref{fig1}.
The cumulants all increase linearly with
the length $L$ of the wire, and the slopes
agree perfectly with Eq.\ (\ref{eq:c123})
[hence the localization length agrees with Eq.\ (\ref{eq:loclen})
and the ratios of cumulants agree with Eq.\ (\ref{eq:cratios})].
The comparison is free of any adjustable parameter.

For contrast,  the upper panel of
Fig.\ \ref{fig2} shows the first three cumulants at
energy $E=0.1$, where the SPS relations (\ref{ccond})
hold and  $C_1= DL/2$  
follows from perturbation theory \cite{Thouless1979}.
The lower panel shows the results at the band edge $E=2$,
which are compared to the predictions
$C_1= 0.7295\, D^{1/3} L$, $C_2=1.602\, D^{1/3} L$, $C_3=0.7801\, D^{1/3} L$
of Ref.\ \cite{finitetimelyap}.

\begin{figure}[t]
\includegraphics[width=\columnwidth]{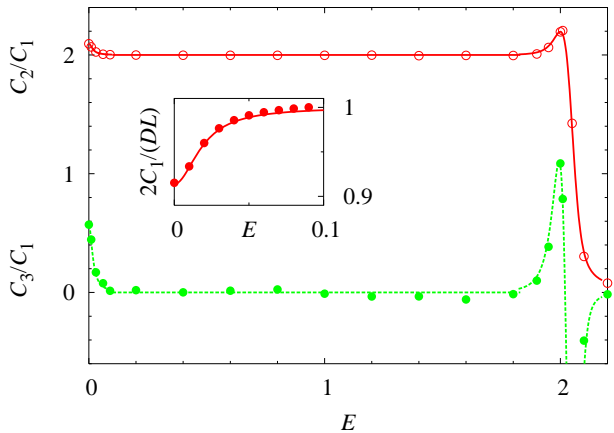}
\caption{Energy dependence of the ratios of cumulants $C_2/C_1$ and $C_3/C_1$.
The inset shows $C_1$  in units of the perturbative result $DL/2$.
The data points are the
result of a numerical simulation of the Anderson model with $D=1/150$.
The curves are the analytical 
predictions of this paper ($E<0.1$), of
perturbation theory \cite{Thouless1979} and 
single-parameter scaling \cite{anderson1980}
($0.1<E<1.8$), and of Ref.\ \cite{finitetimelyap}
($E>1.8$).
}
\label{fig3}
\end{figure}

In Fig.\ \ref{fig3} we show the ratios of cumulants
$C_2/C_1$ and $C_3/C_1$ as a function of
energy. The inset shows $C_1$.
The anomalous region extends up to $E\simeq 10 D$.
Around the band edge, the violations set in for $2-E \lesssim 3 D^{2/3}$.
Again, perfect agreement is found between our analytical theory and 
the results of the numerical simulations.

In summary, we have presented an analytical theory for the 
distribution function $P$ of the dimensionless conductance $g$ in the
localized regime of the Anderson model, Eq.\ (\ref{eq:1}).
The relations (\ref{ccond}) implied by single-parameter scaling theory
for the cumulants $C_n$ of $-\ln g$
are violated not only around the band
edges $|E|=2$, but also 
at the band-center energy $E=0$, where the
correct values are given  by Eq.\ (\ref{eq:cratios}).
Since the random-phase approximation is known to break down in both
cases, our findings reestablish the relevance of this approximation
for single-parameter scaling, which
recently has been contested \cite{Altshuler1,Altshuler2}.

Whether the single-parameter scaling hypothesis itself breaks down
at $E=0$, or just persists in modified form, is
an open question.
The ratios (\ref{eq:cratios}) still imply universal
relations between the cumulants for weak on-site
disorder, i.e., they do not depend on the distribution function of the
random potential.
However, it can be questioned whether this universality 
also extends to additional disorder in the hopping rates, since it is
well known that the extreme case of
purely off-diagonal disorder results in delocalization at $E=0$
\cite{Dyson}.

\end{document}